# An Architecture for Software Engineering Gamification

Oscar Pedreira\*, Félix García, Mario Piattini, Alejandro Cortiñas, and Ana Cerdeira-Pena

**Abstract:** Gamification has been applied in software engineering to improve quality and results by increasing people's motivation and engagement. A systematic mapping has identified research gaps in the field, one of them being the difficulty of creating an integrated gamified environment comprising all the tools of an organization, since most existing gamified tools are custom developments or prototypes. In this paper, we propose a gamification software architecture that allows us to transform the work environment of a software organization into an integrated gamified environment, i.e., the organization can maintain its tools, and the rewards obtained by the users for their actions in different tools will mount up. We developed a gamification engine based on our proposal, and we carried out a case study in which we applied it in a real software development company. The case study shows that the gamification engine has allowed the company to create a gamified workplace by integrating custom-developed tools and off-the-shelf tools such as Redmine, TestLink, or JUnit, with the gamification engine. Two main advantages can be highlighted: (i) our solution allows the organization to maintain its current tools, and (ii) the rewards for actions in any tool accumulate in a centralized gamified environment.

**Key words:** gamification; software engineering; gamification architecture; gamification engine

## 1 Introduction

Gamification is usually defined as the application of game elements and mechanics to non-game activities, in order to improve people's engagement, and motivation, and therefore get better results[1–4]. Successful applications of gamification can be found in many domains, such as marketing, education, or mobile applications, for example. Different types of game mechanics taken from traditional games have been used in gamification[1–3]. The most typical ones are direct rewards in the form of points, badges, or virtual coins the users receive upon successfully completing tasks. Other game mechanics look for exploiting social relations and status, as in the case of levels, leaderboards, or voting. Many gamification applications also make use of feedback systems that provide the users with continuous information on their performance at a given task. The workplace is a very attractive target for gamification. Making work funnier, more motivating and/or more engaging could directly improve the business results of companies and organizations[1–3]. However, gamification itself poses significant challenges. One of them is that we must deeply know the users and their main motivators, and design a gamification solution able to address them and improve the results[5, 6]. In many cases, there are also technical challenges, such as obtaining data from the users' work environment and the tools they use, and to integrate our gamification solution in that work environment.

Though more of a newcomer to the gamification phenomenon, Software Engineering (SE) is no exception. As a matter of fact, the application of gamification in

- Oscar Pedreira, Alejandro Cortiñas, and Ana Cerdeira-Pena are with the Universidade da Coruña, Centro de Investigación CITIC, Laboratorio de Bases de Datos, Facultade de Informática, Elviña, A Coruña, 15071, Spain. E-mail: opedreira@udc.es; alejandro.cortinas@udc.es; acerdeira@udc.es.
- Félix García and Mario Piattini are with the Universidad de Castilla-La Mancha, Grupo Alarcos, Escuela Superior de Informática, Paseo de la Universidad, 4, Ciudad Real, 13071, Spain. E-mail: Felix.Garcia@uclm.es; Mario.Piattini@uclm.es.
\* To whom correspondence should be addressed.
    Manuscript received: 2019-07-22; revised: 2020-01-14; accepted: 2020-01-16

the SE field can have great significance for software process improvement, given that the human factor is the main asset; it is human motivation and engagement that are the keys to success in software projects. In the context of software projects, the engagement of software engineers can be achieved, for instance, by organizing projects as a set of challenges which can be ordered and that need to be fulfilled, and for which some skills, and mainly much collective effort, are required. Software engineers are thus considered as "players" who carry out activities in which they learn new skills, using and combining them to achieve certain challenges, obtain rewards or receive punishments, depending on success or failure, respectively[7]. Gamification can therefore be a very useful instrument to make some environments become fun and attractive (and even addictive). This applies especially to those that include routine and tedious activities, as it is the case with a number of software tasks, such as testing, for example.

Therefore, the field of software engineering has not been unaware of the potential benefits of gamification, and many pieces of research have explored this research line[8]. One of the main gaps identified in existing research derives from the fact that the automation of gamification in software engineering has been achieved so far by developing custom gamified tools (for example, gamified custom tools for requirements analysis and specification, or for software project management). This approach is not feasible in real organizations for two reasons. First, it is common for software companies to use well-known off-the-shelf CASE tools that provide them with a very good functionality level. It may be impossible to incorporate gamification directly into these tools if they are closed products, and it may be also impossible to replace these tools with custom, gamified ones, because of the difficulty of meeting their functionality levels at a reasonable cost. Second, the work environment of most software companies is composed of an ecosystem of tools that support different process areas, such as requirements, project management, development, testing, etc. Even if we could add gamification elements to each of them, it would very difficult to come up with a solution integrating all process areas into a common gamification environment. Therefore, the nature of the work environment of software companies can be a strong barrier for the application of gamification in this domain.

In this paper, we focus on this issue, proposing a software architecture for the gamification of SE environments. As we will see throughout the paper, this software architecture allows any SE organization to incorporate gamification into its workplace without needing to replace any of its current work tools (something that would not be feasible in most cases). Although the architecture will be described in detail in the paper, its main features can be summarized as (1) the core component of our proposal is a gamification engine that will connect to the different work tools of the company through a web service architecture,
(2) this gamification engine is based on an abstract gamification metamodel, and allows the designers to define rules to evaluate and reward the actions carried out by each player in the work tools, and (3) the work tools will communicate each player action to the gamification engine, and those actions will be evaluated according to the gamification rules defined by the designer of the gamified environment. The gamification engine therefore centralizes the logging of the behaviors carried out by each user, along with the evaluation of the game rules that associate the corresponding achievements to those behaviors. The business logic of gamification is thus taken out of the gamified work tools of the organization, and centralized in a gamification engine designed for that purpose. The gamification architecture we present is generic and therefore customizable to any SE organization with its different particular needs and approaches. The software architecture and the gamification engine we propose allow the tools of the environment to be integrated easily through a web service architecture, unifying most of the game mechanics applied in SE in a single tool. This makes for a different approach for gamification in SE, a proposal whose aim is to fulfill the needs of a real software development organization.

In addition to proposing the software architecture and the gamification engine from an abstract point of view, we have implemented a gamification engine based on our proposal, and used it to carry out a case study in a real company. The gamification software architecture we present in this article was developed in a technology transfer project participated by two universities and four software development companies. The gamification engine we developed based on the architecture was used by these four companies, and we were able to carry out a complete case study in one of them. As will see in the description of the case study (Section 5), this gamification engine has allowed us to gamify the complete work environment of a real company, integrating the gamification engine with different work tools, from custom-developed tools of the company, to off-the-shelf and well-known tools such as Redmine, TestLink, and JUnit.

As we will see in the description of the gamification engine, our implementation goes further than just a data centralizer, and enhances the existing tools by providing advanced functionalities, such as the analysis of the graph resulting from interactions between the different participants and sentiment analysis of the texts they introduce into the system. Moreover, it provides a flexible approach that enables the designers to personalize the contents for players according to their profiles, as well as a virtual assistant that can assist the users in knowing how to use the environment. Notice that the purpose of this work is not to show that gamification can improve the results of software engineering companies (that aspect of gamification in SE has already been addressed in previous research works focused on particular ways to gamify particular software process areas), but to present an integral solution for gamifying SE work environments and therefore fill the motivation we have presented above. The rest of the paper is structured as follows. Next section presents related work. Section 3 describes the software architecture for gamification in SE environments. Section 4 presents the gamification engine we have implemented based on the architecture described in Section 3. Section 5 gives details of a case study of the application of the engine for the gamification of a company work environment. Finally, Section 6 presents a discussion on our proposal, and in Section 7, conclusion and future work are set out.

## 2　Related Work

The field of gamification is a vast research area. One of the most significant lines of research in gamification has been the evidence about its usefulness, which was initially evaluated by Hamari et al.[9] by means of a literature review. This study concluded that "gamification does work, but some caveats exist", as most papers report positive results from gamification (with some empirical evidence), but some underlying confounding factors were also present. Gamification in web applications was analyzed in the literature review of Xu[10]. This concluded that gamification was based on superficial game mechanics (point, level, leaderboard, and badges) and that more advanced aspects should be considered, such as social interaction and mobility, by supporting the ubiquitousness of mobile devices, as well as analytics, which must be enhanced.

Much research work has considered the application of gamification in SE, the goal being to improve product quality and project results by increasing people's motivation and engagement[8]. Many software process areas have been considered in previous research work.

For example, Ref. [11] presented a systematic mapping on gamification applied to requirements engineering, where they identified research studies on applying gamification to elicitation, negotiation, and prioritization of software requirements. In Ref. [12], Fernandes et al. proposed a gamified tool, iThink, for requirements management. Reference [13] presented a systematic literature review on the gamification applied to software project management processes. One of their conclusions is that most research works on this topic applied a basic point system reward system, mainly in areas related to integration, resources, and scoping. A good number of gamified tools exist for gamified software project management, such as RedCritter (http://www.redcritter.com), Jira Hero (Atlassian, https://marketplace.atlassian.com/plugins/com.madgnome.jira.plugins.jirachievements), or Scrum Knowsy (http://www.scrumknowsy.com/), all of these with the underlying idea of rewarding users as the project progresses. Several projects, such as Master Branch (https://masterbranch.com/) and CoderWall (https://coderwall.com/), also considered software development in some way, although they are not for a particular software development organization, but rather for communities of developers. Testing has also been considered with proposals, such as HALO[14, 15]. Reference [16] published a systematic mapping on gamification applied to software testing, concluding "the increasing interest for gamification has the potential to lead to positive outcomes". Reference [17] proposed a game called "Code Defenders", where some developers play the role of attackers and introduce errors in the system under testing, while other developers play the role of defenders and have to write test cases that detect those mutant versions of the system. Reference [18] presented an approach for applying gamification to software process improvement, with a focus on small and medium development companies. Reference [19] studied how gamification rules, such as establishing a time limit for development tasks and developers' personal preferences, can affect coding results, such as the working time. Reference [20] proposed a framework for the gamification of enterprise software systems, that is, instead of focusing on the engineering processes that create the product, they focus on the gamifying the system to improve aspects such as user training, acceptance, and usage. Reference [21] addressed an important aspect of gamification, the trade-off between gamification and the participants' privacy.

A more general systematic mapping on gamification in software engineering[8], covering all works addressing the application of gamification to software engineering in any of its areas, found that the adoption of gamification in SE is going more slowly than in other domains, such as marketing, education, health, or banking. This systematic mapping identified two main gaps in the research on gamification in software engineering. One of them is that there is an evident lack of methodological support for the application of gamification elements in software engineering organizations, an issue which was addressed in our previous work[8]. Another important problem in the adoption of gamification in software engineering organizations is the lack of generic architectures and tools for this purpose. Most previous research on gamification in SE has worked with no software support at all, or with custom-developed gamified tools. We believe that this is a very significant impediment. The adoption of software development environments and tools in a real SE company is no small undertaking, and it is by no means cheap. It is not very probable that one of those tools would be changed for another one just because the latter is gamified, since it is highly unlikely that the new tool would provide the same functionality level and set of features as one of the existing, widely-used tools.

The software architecture for gamification proposed in this paper aims to overcome the second set of weaknesses identified in the systematic mapping, by supporting the gamification of an existing SE environment, without replacing any of its current work tools, and centralizing all the gamification logic and additional functionalities, which are described in the next section.

## 3 A Software Architecture for the Gamification of SE Environments

In this section, we present our software architecture for the gamification of SE environments. The proposal has two parts. First, we present the software architecture and its main components: the gamification engine and the software mechanisms to integrate the gamification engine with the organization's Computer-Assisted Software Engineering (CASE) tools. Second, we present the gamification model that has guided the design and implementation of the gamification engine. This gamification model defines the gamification concepts, elements, and techniques supported by the gamification engine, such as (1) behaviors (that represent people's actions in the work environment), (2) achievements (that represent rewards such as points, badges, or resources), and (3) the rules that establish the relationship between behaviors and their corresponding achievements.

### 3.1 Software architecture

The purpose of the architecture is to make the task of gamifying the complete tool suite of a company easier. In order to do this, the business logic related to gamification is moved from the CASE tools to a gamification engine that centralizes and integrates it for all the tools. The basic idea of the architecture is the following: the gamified tools (SE tools covering any software lifecycle activity, such as development, requirements management, project management, or testing, for example) only have to communicate the actions (behaviors) carried out by their users to a central gamification engine. When those behaviors are received in the gamification engine, they are evaluated according to a set of gamification rules defined by the designer of the gamified environment. If a behavior is evaluated as successful according to those rules, the engine will generate the corresponding achievements for the user responsible for that behavior.

Figure 1 shows a high-level view of the architecture. As we can see in the diagram, the gamification engine is the central element of the architecture, since it receives all the behaviors carried out by the software engineers, and evaluates them. The engine provides an integration REST API that allows any other tools to communicate with it. This integration API includes a large list of operations that allow those tools to access all the information from the gamified environment, including those operations for communicating the player's behaviors. Another important part of the architecture is the player's site, which allows players to visualize all the information of the gamified environment, including the user's actions and achievements, and also other gamification elements, such as rankings or progress charts.

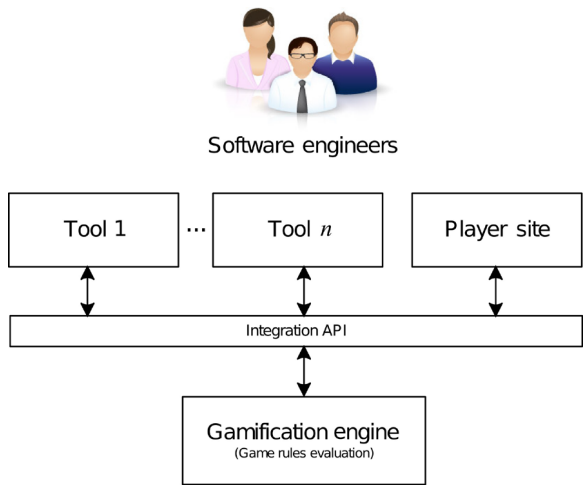

**Fig. 1  High level view of the software architecture for gamification.**

The main advantage of this architecture is that many tools can be included in the same gamified environment. For example, we could gamify tools, such as Jira (https://atlassian.com/software/jira), Eclipse (https://eclipse.org), Redmine (http://www.redmine.org/), or TestLink (http://testlink.org/), the rewards obtained by the players as a consequence of their actions in one of these tools would be added onto the rewards obtained from their actions in any of the other tools. If these tools were gamified separately, it would be difficult to integrate all the rewards obtained by each player. In addition, the logic of gamification would have to be repeated in all of them. However, our gamification engine provides the designer of the gamified environment with generic types of gamification rules that are tool-independent, and which can therefore fit all of them. This design choice simplifies greatly the introduction of gamification in the tools used by the software engineers.

## 3.2  Gamification model

The software architecture and the gamification engine are based on a model composed of three main elements: behaviors, achievements, and game rules. The gamification engine will receive behaviors carried out by the users in their respective tools, and will evaluate these according to the game rules defined by a designer (administrator), to assign the corresponding achievements to those behaviors if the game rules consider them successful.

This model is a central component of our architecture, since it allows the designers of the gamified environments to define behaviors, achievements, and evaluation rules using concepts that are independent of any particular SE work environment we would consider. Although the details of the gamification engine we have implemented are presented in Section 4, some screenshots are included in this section, as they may clarify how the designer can use the concepts of the gamification model in a real case.

In this section we will use a simple guiding example. For the sake of simplicity, let us assume that a generic software development organization wants to gamify its SE environment, focusing on the areas of project management, requirements, and testing. Employee actions receiving awards would include those, such as finishing development tasks, registering requirements in the system, commenting on existing requirements, creating test cases, writing unit tests, or closing the project.

The rest of this section presents the details of the gamification model.

### 3.2.1  Behaviours

Different types of behavior can occur in a software development environment. Instead of trying to identify and model all those particular and specific behaviors, however, we have extracted the features they have in common, and have aggregated them in three types of behavior (summarized in the diagram shown in Fig. 2):

● *Simple behaviors*: This type of behavior is designed for those behaviors where we are only interested in knowing that they have actually happened, as well as who has carried out the behavior, and when there is no need for any other data about the action.

For example, we could define simple behaviors for requirement management actions as being those of registering a new requirement into the system, commenting on an existing requirement to clarify its description, changing its state, or labeling the requirement as completed. These are simple behaviors if we assume that we would not need other data from those actions, apart from the fact that they have happened and who carried them out.

● *Task behaviors*: They are those behaviors in which we are also interested in parameters related to the development and completion of typical tasks in an SE environment, such as the effort, cost, quality, or the completion date. More specifically, the task behaviors currently include the following attributes:

– *Planned completion date*: completion date for the

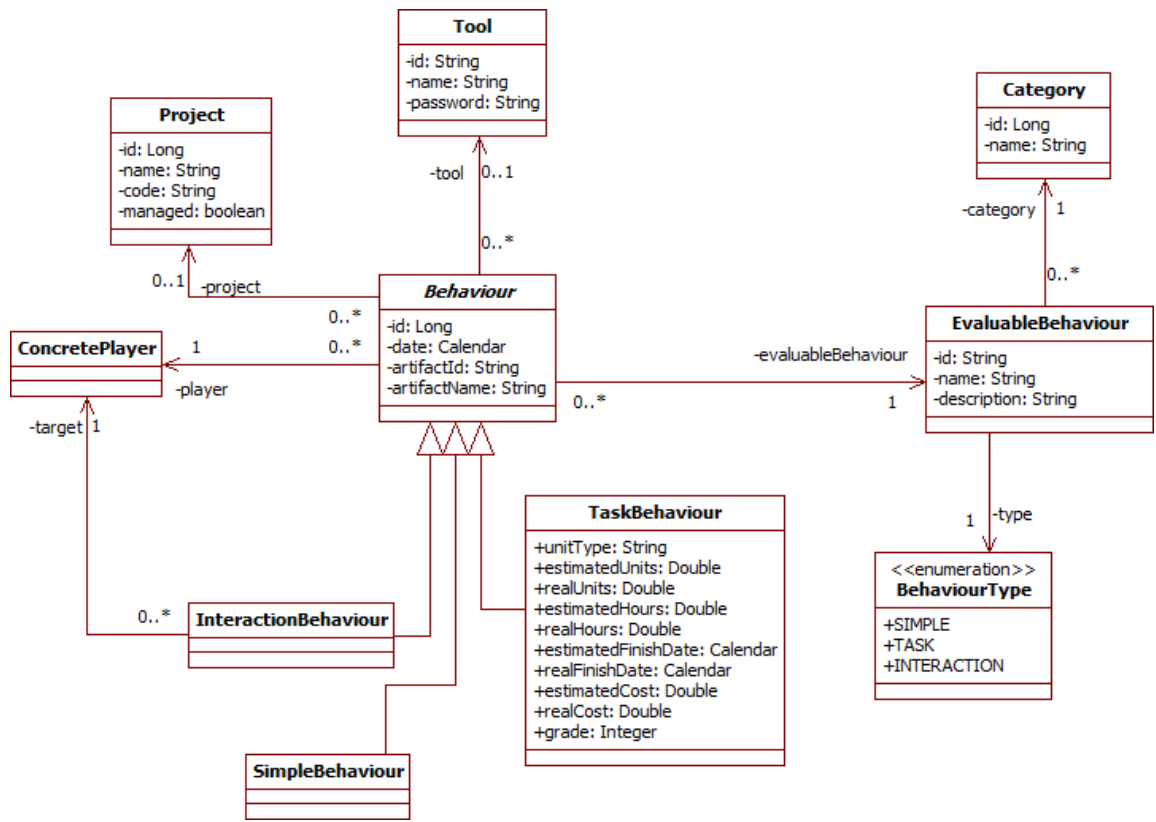

**Fig. 2 Behavior model.**

task in the project plan.

– *Real completion date*: date on which the task was actually completed.

– *Estimated effort*: estimated effort, in hours, to complete the task.

– *Real effort*: real number of hours needed to complete the task.

– *Estimated work units*: work units refer to tangible results of the task, such as lines of code, classes, or requirements, for example. This attribute corresponds to the estimated number of work units for the task, if this has been estimated.

– *Real work units*: real number of work units completed during the task.

– *Unit type*: the name of the work units that are being considered.

– *Grade*: this attribute, which takes values between 0 and 100, allows us to take into account the quality of the results obtained during the task.

As we will see later, these attributes of a task behavior can be used in the definition of the gamification rules. For example, we could reward the finishing of a task only if it has been completed by the planned completion date, with the estimated effort, and with a given quality level. None of these attributes is mandatory, so we can use just those ones that are of interest in each particular case.

Task behaviors are designed mainly for those tasks that would appear in a project plan, or in a product backlog, for example. The most obvious example for task behaviors is development tasks. That is, when a developer marks a task as completed, the system would notify that action to the gamification engine, indicating the estimated and real dates and effort. However, task behaviors may also apply to other actions.

• *Interaction behaviors*: They represent actions in which two people have collaborated in some way. This type of behaviors is concerned with rewarding the collaboration in the workplace. For example, we could use it to record that two people have interacted because one of them has created a task and has assigned it to the other, or because one of them has registered a requirement in the system, and the other person has commented on that requirement.

As we will see in the next section, these classes of behaviors will also allow us to derive an interaction graph from which important information can be extracted, such as the interaction network of each user, relevant users that act as hubs or links, and the existence of communities that can be automatically identified from this information.

Figure 2 shows a class diagram summarizing the behavior types. As we can see in the diagram, the model also considers maintaining who has carried out the task, the tool from which the behavior was received, the project in which it has been carried out, and the date and hour on which the action has taken place. These attributes allow the gamification engine to keep a persistent log of all the actions carried out by team members in each project, which is a valuable information.

Notice also that all behavior types include two more attributes: *artifactId* and *artifactName*. Most tasks in an SE environment give as a result a project artifact, such as a document, or a task in the project plan, for example. These attributes allow us to include in each behavior the identifier and name of the resulting artifact. For example, in the behavior "Task completed", we could indicate the identifier and name of the task. As we will see in the presentation of the gamification rules, the attributes can also be used in the definition of the rules, as well as in the messages that will be shown to the user when receiving a reward. These three types of behavior cover most actions that could take place in an SE development environment. Although the model currently considers these types of behaviors, it could be easily extended to support new ones, if we detected a kind of action that does not fit in to these three types.

When configuring the gamified environment, the administrator will start by defining the behaviors that are subject to being evaluated and rewarded. For each of them, the administrator will have to indicate for each behavior only its identifier (a string), its type (simple, task, or interaction behavior), its name, its description, and its category. The identifier is a key point, since it will be used by the gamified tools when communicating behaviors to indicate what action they are communicating.

*Example*: Figure 3 shows a screenshot of the behavior definition screen in the gamification engine we have implemented. In this example, we have defined just four behaviors: create a task (GSE_CREATE_TASK), complete a task (GSE_TASK_COMPLETED), detect an error (GSE_ERROR_DETECTED), and comment on a project requirement (GSE_COMMENT_REQ).

#### 3.2.2 Achievements

When the rules of the game determine that a user has successfully completed a behavior, the system will reward that user with an achievement. The model has been designed to provide a flexible range of achievements. Three classes of achievements are currently supported:

• *Points* (also called experience points): They are the basic reward mechanism, with a role analogous to what this type of achievement has in classic games. The number of points is a measure of the amount of successful behaviors completed by each user. In addition, the experience points also determine the *level* of the player.

The environment designer could even define more than one type of points (in order to distinguish between clearly different groups of behaviors). However, one of them must be used as the basis for computing the level of each player.

• *Badges*: They are a classical achievement type in gamification. Badges are usually granted when a significant milestone in the gamified environment is reached.

The designer of the gamified environment can define as many badges as needed. For example, we could grant a badge on a developer's first 100K line of code, or

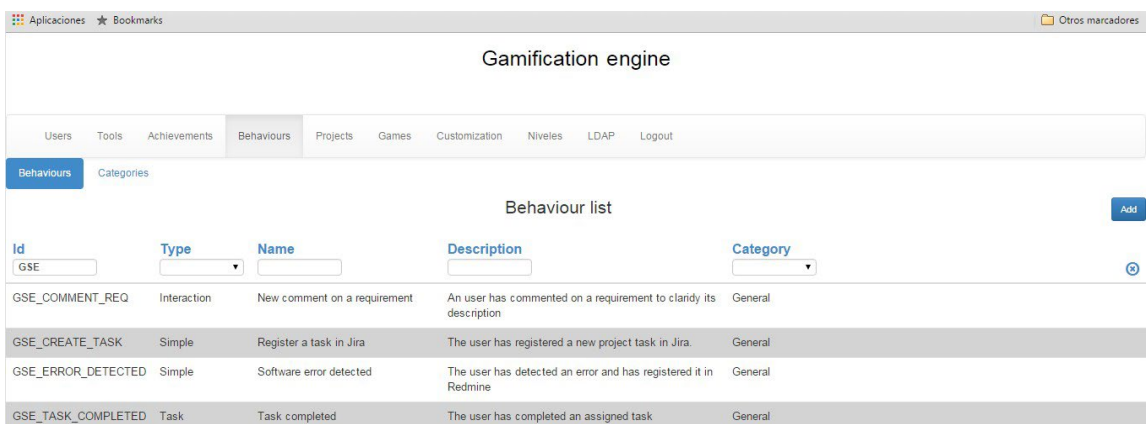

**Fig. 3　Screenshot of the behavior definition screen.**

Oscar Pedreira et al.: *An Architecture for Software Engineering Gamification*     783

establish badges for the best analyst, best developer, and best tester of the month. Other badges could be created, depending on the design of the gamified environment.

• *Resources*: They meant to represent real-world rewards for the players. For example, resources could be used to reward the players with physical gifts, or time packages they can devote to personal projects or training, for example.

This set of achievements will allow us to apply the most typical game mechanics used in gamification. Experience points, badges, and resources are all direct rewards, but we can also use them to implement levels, leaderboards, social status, and even quests.

The diagram shown in Fig. 4 summarizes the achievement classes currently considered in our gamification model. The model allows the environment designer to define as many achievement types as needed, each of them belonging to one of the achievement classes we have just established. That is, although the gamification model currently provides the three types of achievement we have presented, it allows the designer of the gamified environment to define new types of achievements, like currencies.

*Levels*: although levels are not a particular class of achievement, they are directly derived from the experience points of the players through an exponential function that can be customized by the environment administrator,

$$f(l) = a \times b^{l \times c},$$

where $l$ is the level, and $f(l)$ returns the number of experience points necessary to achieve level $l$. For example, with values $a = 1$, $b = 1.4$, and $c = 2$, the number of points necessary to achieve the first nine levels are shown in Table 1.

In this way, the difficulty of getting to the next level is completely customizable. It can be made linear, or exponential, as in our example, making it increasingly difficult to get to the next level.

**Table 1 Exponential function for levels.**

| Level | Number of points | Level | Number of points | Level | Number of points |
|---|---|---|---|---|---|
| 1 | 1 | 4 | 14 | 7 | 111 |
| 2 | 3 | 5 | 28 | 8 | 217 |
| 3 | 7 | 6 | 56 | 9 | 426 |

### 3.2.3 Gamification rules

The link between the user's behaviors and the achievements is established by the gamification rules. The model provides a gamification rule system that allows the environment designer to define a complete set of rules in a flexible way. This is the most important component of the model, since it removes the logic of gamification from the gamified tools, and it allows centralizing it in a gamification engine.

A game rule maps behaviors to achievements. Each rule has a source type of behavior and many target types of achievement. Every time a behavior from the source type is received at the engine, all game rules with that source type of behavior are activated and evaluated by the gamification engine. Each rule is associated to its types of achievement through an achievement modifier, which represents the condition that uses the behavior's attributes to define the criteria determining whether the achievement is granted or not.

*Example*: In the example we are using for presentation of the gamification model, the organization could be interested in defining a rule for the behavior "Task completed", which is a task behavior. On receiving such a behavior, we would like to reward the user in different ways depending on whether or not the task has been completed within the parameters of estimated effort. The definition of such a rule is shown in Table 2.

As we can see in the example shown in Table 2, the rule "Task completion" is activated when a "Task completed" behavior is received, and three possible achievements are evaluated. In the first one, if the user has completed the task with an effort less than the estimated one, he or she is rewarded with as many experience points as the effort estimation of the task. In the second achievement, if the real effort is greater than or equal to that estimated, the user is rewarded with a number of points equal to what was estimated, minus a penalization for the number of hours he/she has exceeded the estimation. Finally, in the third achievement, if the user has completed the task in less than half the estimated effort, he is rewarded with an extra badge of "Star performer". It is important to notice that all the conditions and modifiers used in this example can be specified in the gamification engine using the behaviors attributes.

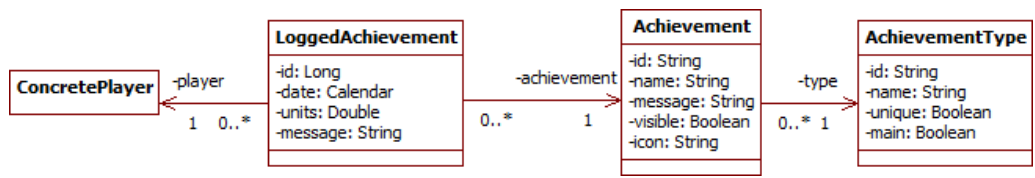

Fig. 4 Achievement model.

**Table 2 Example of a gamification rule. The rule name is "Task completion" and the source type of behavior is "Task completed".**

| Achievement | Condition | Achievement result | Modifier |
|---|---|---|---|
| 1 | realEffort<estimatedEffort | Experience points | estimatedEffort |
| 2 | realEffort ≥ estimatedEffort | Experience points | estimatedEffort −(realEffort-estimatedEffort) |
| 3 | realEffort<(estimatedEffort/2) | Star performer badge | – |

The gamification engine could now receive "Task completed" behaviors from any tool, such as Jira, or Redmine, for example, which would communicate those behaviors with the real attributes of how a user has completed a task in that tool. Let us look now at what would happen in the following three cases:

• *Case 1*: John completes a task with 20 estimated hours in just 18. In this case, he is rewarded with the Achievement 1; that is, 20 experience points.

• *Case 2*: John completes that task in 22 hours. In this case he receives the Achievement 2, that is, 18 experience points $(20 - (22 - 20))$.

• *Case 3*: John completes the same task in just 8 hours. In this case, John will receive two Achievements, the first and the third. For the first one he receives 20 experience points, and for the third one he receives a "Star performer" badge, since he has completed the task in less than half the estimated time.

Figure 5 shows a screenshot of the rule definition screen in our gamification engine implementation, with the same example we have just presented. As we can see in the screenshot, this rule, "Task completion", will be activated when a "Task completed" behavior is received, with three possible achievements for the user who has completed the task. In Fig. 5, we can also see that, in addition to the definition of the conditions for each achievement, the administrator of the gamified environment can introduce messages that will be shown to the user who completed the task, when obtaining each of the achievements. Notice that the messages like "*Congrats! You've completed a task! (Task #id, #name)*" can also use the attributes of the behavior that has been evaluated. In this example, *#id* and *#name* correspond to the attributes *artifactId* and *artifactName* of the task behavior. When evaluating a particular behavior, the message template would be transformed into a real message, such as "*Congrats! You've completed a task! (Task 45, User authentication)*". These messages can be shown on the player's site or in the tools that have communicated the behavior to the gamification engine. The example shown in Fig. 5 demonstrates that the designer also has the option of awarding a given achievement to a type of behavior only the first time that a behavior of that type is evaluated. This would allow us to define a rule that, for example, awards a "*First task completed!*" badge to a player only the first time he/she carries out a task in the system. It is worth highlighting that the conditions of the rules are established through the graphical interface of the engine, that is, without modifying its source code.

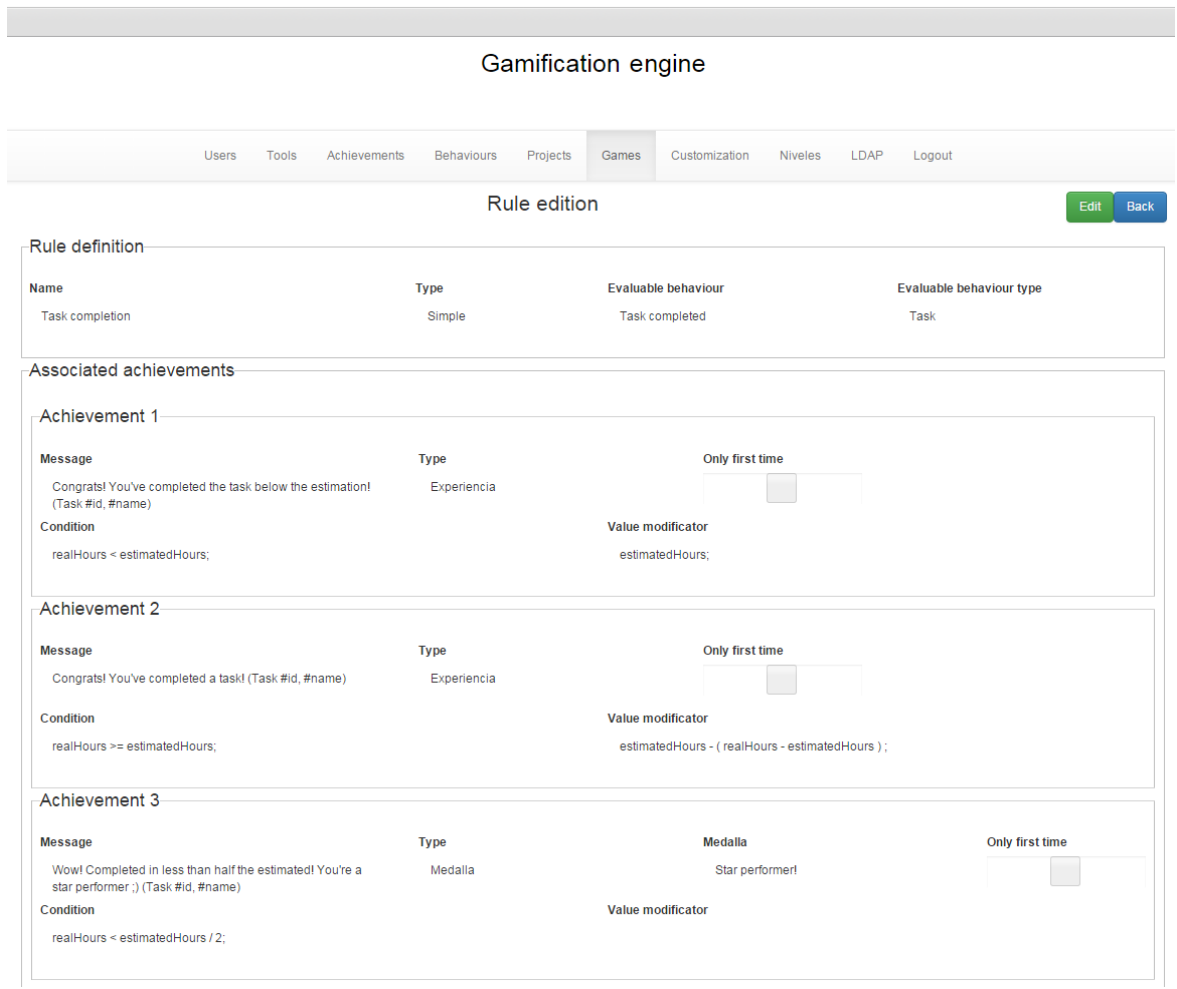

**Fig. 5 Screenshot of the rule definition screen.**

Although this example is simple, it shows the flexibility of the rule system of the gamification model. As we have just seen, the designer of the gamified environment can establish any condition on the received behaviors, and can also use its attributes when awarding achievements. This provides us with a high degree of flexibility in rule definition. We should also point out that the tool in which the user carries out the behavior knows nothing about how the action is gamified; it simply has to communicate it to the engine. This allows us to integrate and therefore to gamify as many heterogeneous SE tools as required. The example we have just shown considers the simplest type of rule supported by the engine. Actually, we distinguish between three types of rules:

• *Simple rules*: They are gamification rules that evaluate just the condition of each achievement on the received behaviors, determining if an achievement must be awarded to the player.

• *Repetitive rules*: These rules award the achievements only when the conditions are evaluated successfully a given number of times; in other words, a number of behaviors that fulfill the required condition were received. Besides, it can be specified that the behaviors must be received within a closed period of time, defined by start and end dates.

• *Interval repetitive rules*: These are also repetitive rules, but instead of defining start and end date, a generic interval of time (i.e., week, month) is selected, so a number of behaviors that fulfill the condition must be received within this period.

These types of rules allow us to reward behaviors not just when they happen, but when they happen repeatedly in time. For example, we could reward a developer for completing one hundred tasks, or for completing those one hundred tasks in a month.

Figure 6 shows a class diagram summarizing the design of the gamification rules in the model. Since a complete gamified environment can have a large set of rules; these can be grouped into games. In this context, therefore, a game is defined as a set of related rules. The designer can even configure that only some particular games are played in a project.

Notice that although the examples we have used in the description of the rules involved mainly task completion time, the types of rules we have considered allow us to reward behaviors more than just finishing on time or in cost. In addition, that the *TaskBehavior* type of behavior includes a grade attribute, intended to reflect the quality of the work.

## 4 Gamification Engine for SE Environments

In addition to proposing the generic software architecture and gamification model presented in the previous section, we have also implemented a gamification engine based on them. This implementation has allowed us to carry out a case study on the gamification of the work environment of a real software organization using our proposal. In addition, we have incorporated to our gamification engine functionalities that can be of interest for a real organization, and that go beyond the gamification architecture and model we have presented. These functionalities include a social network for the players, messaging, system notifications, challenges between players, and a virtual assistant based on dialog-generation technologies. It also contributes tools for the analysis of the activities carried out by the users; in particular, there is a tool for analyzing the interaction of players, and community detection based on the interactions of the users in the workplace, along with a sentiment analysis module. This module enables detection of positive and negative polarities in the texts

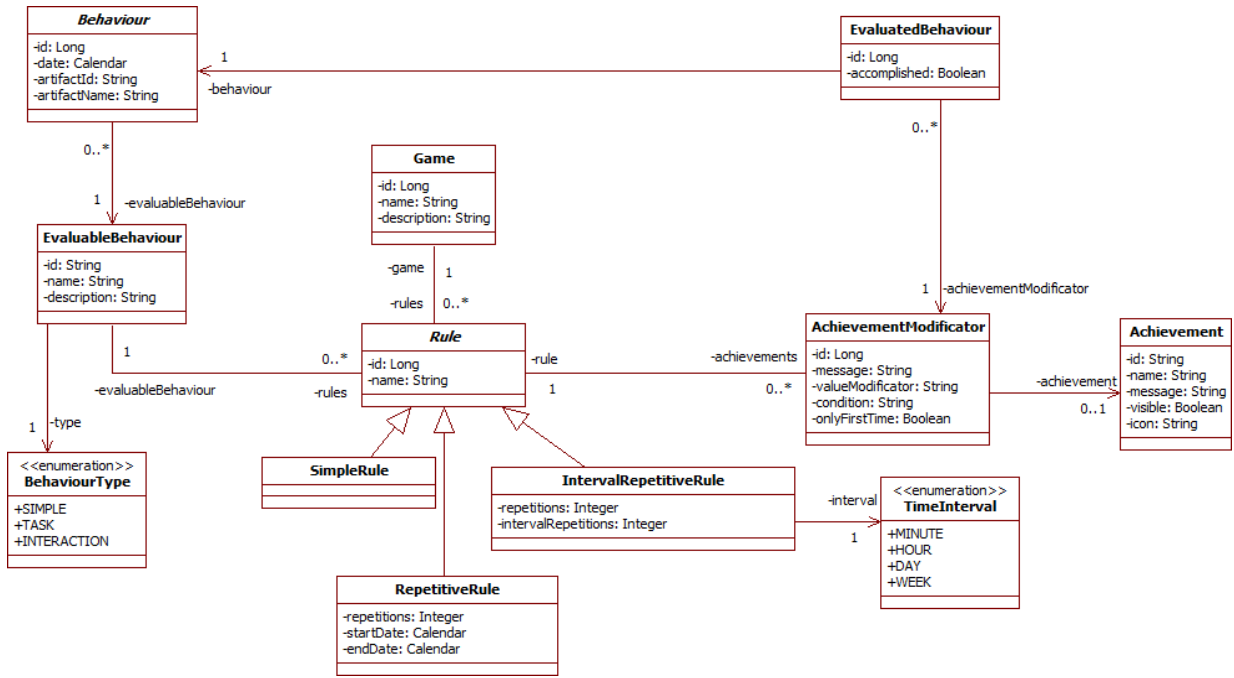

Fig. 6 Rules of the gamified environment.

introduced by the players.

In this section we present the details of the implementation of the gamification engine, how it supports the software architecture and gamification model presented in the previous section, and the additional advanced functionalities we have added.

### 4.1 System architecture and design

The gamification engine has been designed following a three-layer architecture (see Fig. 7). The first layer is devoted to data persistence, and has been implemented as a relational database in PostgreSQL. The engine model contains the data access layer and the business logic we have described, comprising the management of users, behaviors, achievements, and gamification rules. A third layer provides two different interfaces.

• The administrator of the gamified environment accesses the configuration of the engine through a web application that provides an interface from which the administrator can manage everything: users, tool credentials, behaviors, achievements, game rules, etc.

• REST API provides a complete interface for all the tools of the gamified environment. This interface provides those tools with a large set of operations that allows them to access all the information in the gamified environment, and not just the communication of behaviors. For example, the player's site does not have its own database, since it accesses all the information stored in the engine through the API.

The engine has been developed in the Java EE platform, using technologies such as Hibernate, Spring, Spring MVC, and AngularJS.

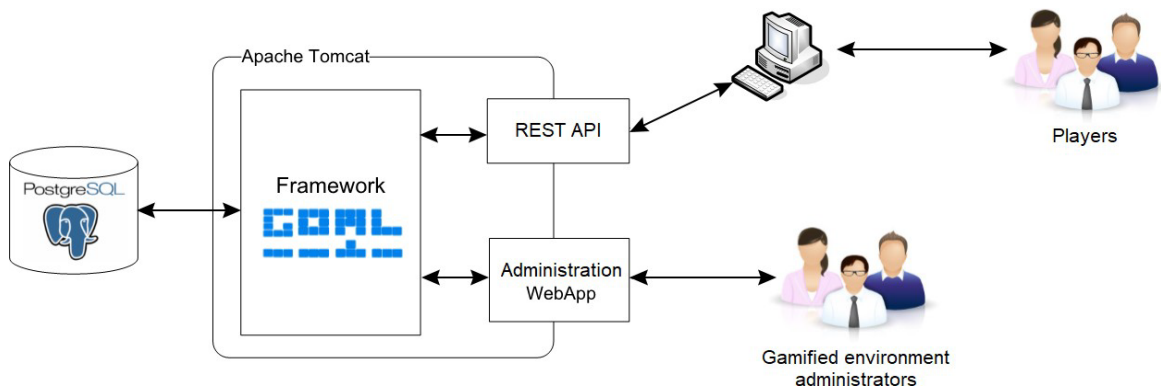

Fig. 7 Engine architecture and design.

## 4.2 Integration

The REST API provides an interface for the rest of the tools in the gamified environment. By implementing this interface as a REST web service, we ensure that the platform or technology will not be an impediment for integrating any tool into the engine. In addition, since many tools that could be integrated into the engine have been developed in Java, we have created a client library for the REST API; this simplifies its use.

The gamified tools cannot communicate freely with the engine. If a tool has to communicate behaviors, it must be registered in the engine with a tool ID and a password that will be used in every transaction.

So far, we have focused on the data sent by the working tools to the gamification engine for the purpose of registering the player's actions and evaluating them according to the gamification rules. However, data flow in the opposite direction is also possible, since the REST API allows all gamified tools to access all the data related to players, their actions, rewards and, in general, all gamification information (even the rules of the game). In this way, the gamified tools could also show the results of the gamification live to the players. For example, a programmer could see the result of a just completed development task in the Integrated Development Enviroment (IDE) he/she is using.

## 4.3 Other functionalities

The engine completes the basic gamification model we have described with other functionalities present in classic games that are also used in gamified applications.

● *Social networks*: Most collaborative games allow players to communicate with their friends, or to even have an explicit social network. The engine supports this concept by providing a social network among the players, who can explicitly create friendship relationships.

This allows us to, for example, show different rankings to the users, as rankings comparing their results with that of the rest of the players, or as a ranking comparing the results obtained by one user with those of their friends. Although the gamification engine currently provides its own social network, the data about the player's relationships could be obtained from an external social network if the company is already using one.

● *Messaging*: This is a feature present in most collaborative games, allowing the players to communicate with their peers instantly.

● *Profile information and rankings*: One of the important game elements used in gamification is continuous feedback on the actions of a user; that is, the users can immediately see the results of their actions in the games. The engine covers this need in two ways. First, the achievements assigned to each received behavior are returned to the application that communicated the behavior, so they can be immediately shown to the user. Secondly, the engine provides all the tools with the whole set of information making up the user profile (personal data, level, and achievements obtained to date), also giving rankings that allow users to compare their performance in the gamified environment with the performance of the rest of the users (all users, or only their friends).

● *Quests*: Quests allow users to challenge other users to achieve a certain goal in a given period. That goal can be expressed as a certain number of points or badges of a given type.

## 4.4 Support of game mechanics and elements

A popular question in gamification is what game mechanics we can apply in gamified environments in order to foster motivation and engagement in the users. In our previous systematic mapping about gamification in software engineering[8], the game mechanics and elements which have been considered previously were identified. Table 3 shows that list of game mechanics, and how they are supported in our gamification engine. As we can see in Table 3, only one, namely "betting", is not currently supported.

## 4.5 Player's site

A fundamental part of the gamification engine is the player's site, which allows them to see all their activities in the gamified environment. Figure 8 displays a screenshot of the home page of the player's site in a real setting of the engine (real logos have been removed from the head of the page). This application allows the players to see all the information about their activity in the gamified environment. The home page shows them their profile information, the experience points they have accumulated, the level, the percentage of points obtained until the next level is reached, a chart for experience points, a list of the badges obtained, and two rankings, one of them considering all the players, and the other one taking into account only the players immediately above and below the player. The site also allows the players to access other information, such as a map with their locations, the projects they are involved in, social networks ("Friends" option in the menu), messages, notifications, challenges, and access to the virtual assistant.

**Table 3 Game mechanics support in the engine.**

| Game mechanic | Support in the gamification engine |
|---|---|
| Points | Experience point |
|  | Configurable points (currencies, karma, etc.) |
| Badges | Badges |
| Levels | Computed from experience points, and configurable |
| Continuous feedback | Player's site provides real-time data on achievements |
| Game dialogs | Virtual assistant |
| Quests | Challenges created by users |
| Rankings | Presented on the player's site |
| Social network | Supported in the engine, and shown on the player's site |
| Voting | Can be supported through task behaviors |
| Betting | Not supported |

The players can thus access all the information of the gamified environment in a single place. Of course, this does not prevent us from showing information about rewards in the gamified tools.

## 4.6　Advanced functionalities

In this section, we describe other advanced functionalities of the engine, such as the support for customization and a virtual assistant that can provide help to the users using natural language; we also give a description of functionalities for sentiment analysis and interaction network analysis.

### 4.6.1　Customization

This module supports the inclusion of personalization rules in the system. The administrator can define variables with an associated condition. That condition is an arithmetic-logic predicate that can use the attributes of the user's profile. In this way, when evaluated for a particular user, each variable will return a true/false value. This would allow us to show some parts of the environment to one group of users, and not to others.

The expression elements we can currently include in the definition of customization rules are listed in Table 4.

*Example*: Using the customization variables presented, we could define the following customization rules (shown in Table 5): the first rule would tell us to search for, and suggest friends for, those people who have not reached Level 5, and who have fewer than 20 friends. In the second case, the rule would tell us to show a system tour for rookies, that is, for users that have not reached even Level 2.

**Table 4　List of variables available for customization rules.**

| Expression | Meaning |
|---|---|
| Date(date) | A given date |
| Date | Today's date |
| firstBehaviorDate | Date of the first behavior of the player |
| Points | Accumulated points of the player |
| Level | Level of the player |
| Followers | In-degree of the player in the interaction graph |
| Following | Out-degree of the player in the interaction graph |
| Polarity | Average sentiment polarity of texts of the last five days (takes values between $-1$ and 1). |

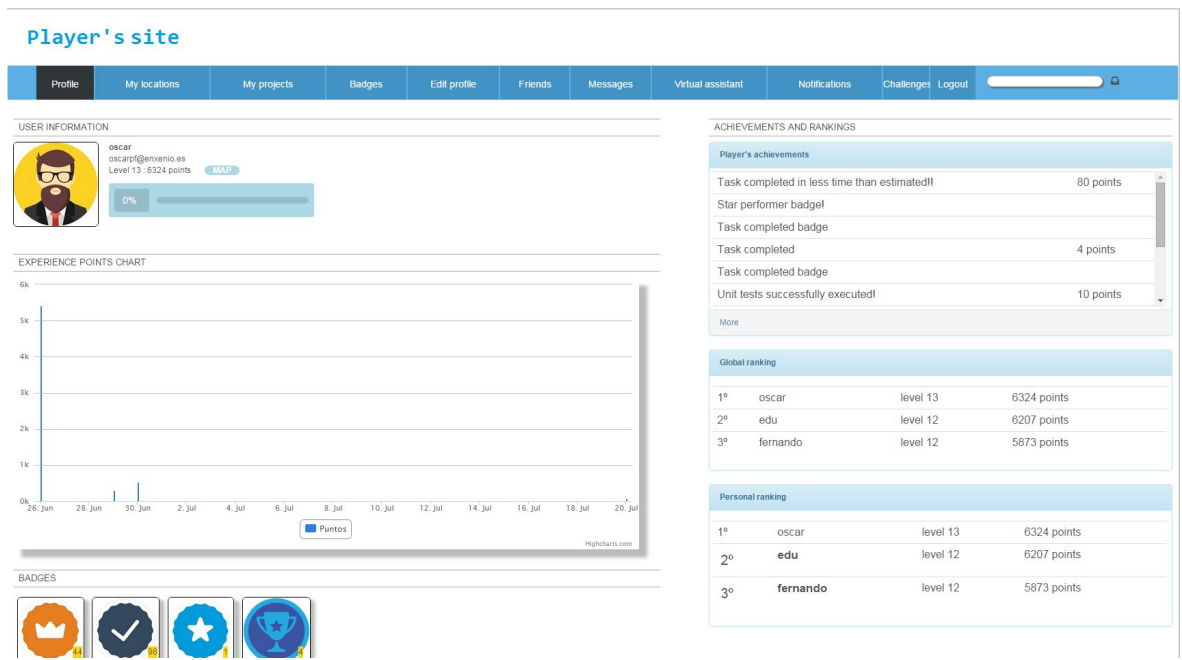

**Fig. 8　Screenshot of the player's site home page.**

Table 5 Example of customization rules.

| Customization variable | Condition |
| --- | --- |
| SUGGEST_FRIENDS | Level <5 & Following <20 |
| SYSTEM_TOUR | Level <2 |

### 4.6.2 Virtual assistant

We have included a virtual assistant in the engine, based on natural language generation technologies. Its purpose is to provide help about the work environment, not with static pages, but with generated dialogues, which is closer to what happens in real videogames.

This assistant has been implemented as a chatbot using Alice (i.e., Artificial Linguistic Internet Computer Entity)[22], a natural language chatterbot that generates dialogues by applying heuristic pattern-matching rules to the texts introduced by people. The dialogues the virtual assistant can process are expressed in AIML (i.e., Artificial Intelligence Mark-up Language)[23] files, based on categories, patterns, and templates. We do not provide more details on how to write the AIML files, since this would be outside the scope of this paper.

Leaving the internal details of dialogue writing aside, this functionality allows us to provide the players with an interface with the system in natural language. This virtual assistant can give, for example, information about the different tools and processes the organization applies in the SE environment, as well as about how to progress in the gamified environment.

### 4.6.3 Interaction network analysis

When used in an organization with many people working together with different gamified tools, the gamification engine will receive and generate a lot of information about the behavior of the users, the results they achieve with their work, and how they interact and collaborate. This is especially important in software projects, which has motivated us to incorporate additional components into the engine, aiming to provide the administrator with tools to analyze information that would be difficult to obtain without the use of the engine.

An interaction graph can be derived from the set of interaction behaviors received by the engine. The set of users of the gamified environment is the set of nodes of the graph, and the set of interaction behaviors is the set of edges. Notice that the edges are labeled, since the administrator can define different types of interaction behavior. For example, we could create an interaction behavior *Collaborate*, with the semantics registering that two users have collaborated in the completion of a task; that is, they have worked together to carry it out. In addition, we could create a second type of interaction behavior *Helps* to represent that a user has helped another user in his/her work (by providing information or knowledge, for example).

This interaction graph is a valuable information asset to see the behavior of all the members of an organization and how they interact. The engine thus does not only provide a way of gamifying a workplace, but also a system with which to gather and analyze relevant information on the organization's dynamics. This graph may allow us to identify flows of information in the company, relevant users that act as hubs (that is, they are central to the connection of many people), and even to detect communities inferred from interaction information.

Figure 9 shows a screenshot of the community detection module. In this example we can see that two communities have been identified (the nodes of each community are shown in purple and green, respectively). This module provides different algorithms for the detection of communities, namely Edmonds-Karp[24], Girvan-Newman[25], Tarjan[26], and Louvain[27]. The example seen in Fig. 9 has been created from a sample of the interaction graph obtained in the application case study we present in the next section.

### 4.6.4 Sentiment analysis

The social network included in the engine will contain a lot of information as messages exchanged between the users. This happens with messaging or chatting, or with the gamification engine, as is the case of a chat with the virtual assistant. Between users, this information will usually reflect more personal communications, with a different register from those texts written by the users in work tools, such as Redmine, for example. These social network texts can therefore show a more biased content, and so be subjected to sentiment analysis.

The goal of a sentiment analysis classifier is to take a given text and classify the polarity of that text automatically into positive, negative, or neutral[28]. The engine includes a sentiment analysis module that allows the texts written by the users to be analyzed; it can then tell us their polarity. This module has been implemented using a machine learning approach[28] with Support

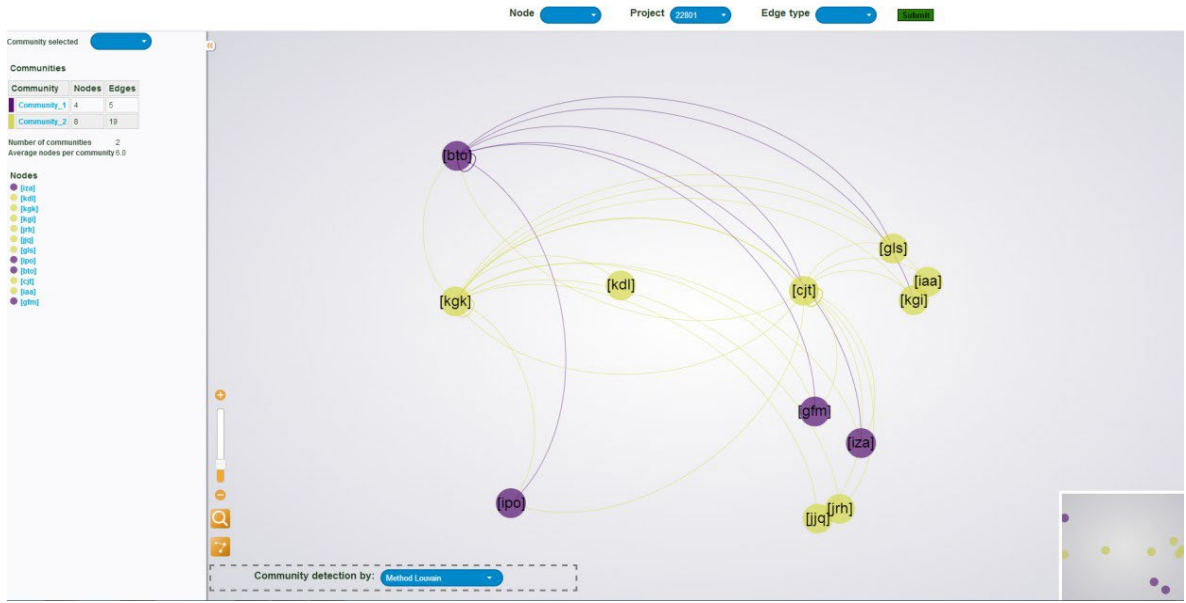

**Fig. 9 Screenshot of the visualization of community detection.**

Vector Machines (SVM), which is a supervised machine learning method. The training set (a collection of texts that have already been classified by a human) for this classifier includes an ad-hoc dataset created for the project from the database of work tools of a real company. It could, nonetheless, be easily enriched with data from other companies. The classifier currently reports text polarity with a precision of around 80%, which is consistent with the current results of the state of the art in this field.

This component allows us to analyze every text introduced by the user, and to detect situations in which a user clearly shows a negative trend in his/her latest messages. These data complement the user profiles with information that goes beyond their personal data, and beyond the log of their actions at work. The output of this component is stored in the database of the gamification engine as a list of classified texts for each author. In addition, as we said earlier, the output of the sentiment analysis component can be used in the personalization component as well. This would allow us to detect users with a negative tendency, show them personalized contents, or suggest that they talk to their friends (who are also kept in the gamification engine). The operations for sentiment analysis are also accessible from the gamified tools through the integration API.

## 5　Case Study: Application of the Gamification Engine in a Real Company

In this section, we present a case study on the application of our proposal in a real company. We have used the case-study method, following the template for case studies presented in Ref. [29], and the guidelines proposed in Ref. [30]. We present the background, design, subjects and analysis units, field procedure and data collection, intervention, and an analysis of the results obtained in the case study.

### 5.1　Description of the organization and its tool suite

The case study took place in a small software development company. We will refer to this company as SC throughout the paper. SC focuses mainly on software development, and it currently offers products sold as off-the-shelf packages, as well as custom development services for its customers. The firm currently employees 25 people, 18 of them devoted to software development. Its areas of expertise include software for business management, education, digital contents, electronic commerce, and geographic information systems.

SC has certified quality management systems for software development, under ISO 15504/ISO 12207 (SPICE)[31, 32], and information security management, under ISO 27001[33]. It also has experience on software product certification under ISO 25 000[34].

The software engineering environment of SC comprises many tools. SC has, importantly, developed a custom tool for project and requirements management, which we will call SC-Manage. This tool allows project managers to register the project plans and requirement books, to assign tasks to people, and to perform project monitoring. The tool is fed with the effort reports registered by the team members. It therefore has complete information about the project management and the requirements of the project. SC also uses complementary tools, such as Redmine for issue management, TestLink for test plans, JUnit for unit testing, and SVN/GIT for version control.

## 5.2 Design

According to the approach presented by Ref. [35], the design type of the case study is single case — holistic, since we have focused on the single case of SC. The object of the case study is the gamification engine we have presented in Section 4. The main research question of the case study was: is the gamification engine a suitable tool for creating integrated and multi-tool gamified software engineering environments? Table 6 presents the Main Research Questions (MRQ) and Secondary Research Questions (SRQ) of the case study. Our main research question directly addresses the motivation of this work, presented in the introduction of the paper. That is, with this case study we want to validate if our gamification engine would allow us to implement a gamified work environment in real software companies. Therefore, we want to validate that (1) it must be able to integrate and accommodate a wide range of CASE tools, either off-the-shelf or custom developed, without needing to replace them (SRQ 1); (2) the set of behaviors, achievements, and rules provided by the gamification engine must meet the needs of the designer of the gamified environment; and (3) the effort required to integrate the organization's CASE tools in our framework must be reasonable, that is, it should be by far smaller than that the effort required to replace the organization's tools or to develop a custom gamification software.

As it is described in the rest of this section: (1) the case study was conducted in a real company that works with both well-known off-the-shelf (such as TestLink, Redmine, and JUnit) and custom tools (such as the one used in project management); (2) the implemented gamified environment integrates all these tools and makes use of a wide range of game mechanics similar to those already used in previous works on gamification in SE; and (3) the effort required to implement this environment can be considered really small if compared to the effort required to replace any of the company's tools or to develop a gamification software from scratch.

## 5.3 Subjects and analysis units

The company SC has already been presented in Section 5.1. Along with other companies, SC participated in a broader research project focused on the application of gamification in software engineering environments. The analysis unit of the case study is the gamification engine, including the integration API and the player's site.

## 5.4 Field procedure and data collection

The execution of the case study comprised the activities of scope and solution definition, analysis and design of the gamified environment, and development of the gamified platform. The authors of this work took part in the execution of the case study, providing support in the design of the gamified environment, the use of the gamification engine, and the integration of the SE tools. Data related to the design and development of the gamified environment were kept in the form of documents. Data were also obtained from direct interviews with the team members. Finally, data about development efforts also came from the records of the project management tools of SC.

## 5.5 Intervention

This subsection summarizes the main aspects of the execution of each phase of the business case.

### 5.5.1 Scope and solution definition

As we explained in the background to the case study, the tool suite of SC is composed of many tools, as shown in Table 7. The most important one is SC-Manage, which supports project management and requirements management, and which is a custom development of SC. However, SC also uses Redmine for issue management, TestLink for test plans, and JUnit for unit testing.

In this project, the goal of SC was not to gamify just one single tool, but to include all of them in the same gamified environment. This meant that all the tools we have mentioned were taken into account within the scope of the case study. Table 7 shows the list of behaviors considered in the design of the gamified environment, along with the particular tool where the employees carry out those behaviors.

**Table 6 Research questions of the case study.**

| Research question | Description |
|---|---|
| MRQ | Is the gamification engine a suitable tool for creating integrated and multi-tool gamified software engineering environments? |
| SRQ 1 | Is it feasible to integrate different SE tools, including COTS from different providers, in a single and centralized gamification environment using the gamification engine? |
| SRQ 2 | Does the gamification model of the engine (behaviors, achievements, and rules) support a real gamified environment? |
| SRQ 3 | Does the engine allow us to create a gamified environment with a reasonable development effort? |

Table 7  List of behaviors communicated from each tool.

| Tool | Behavior |
|---|---|
| SC-Manage | Create task |
| | Assign a task to people |
| | Report task effort |
| | Complete a task |
| | Open requirements book |
| | Create requirement section |
| | Register requirement |
| | Update requirement state |
| | Add attachment to requirement |
| | Add attachment to requirement |
| | Close requirements book |
| Redmine | Serious bug in development |
| | Serious bug in production |
| | Minor bug in development |
| | Minor bug in production |
| | Close issue |
| TestLink | Create test plan |
| | Create a test case |
| JUnit | Run unit tests |

All of the behaviors included in the list were simple behaviors, except for "Report task effort", "Complete a task", and "Run unit tests", which were considered task behaviors. In the first two behaviors, the use of the attributes is directly related to the task. In the effort report, only the "Real effort" attribute is used, with the value of the reported work hours. For "Complete a task", all the attributes of the behavior have been used, as in the examples we have presented in the description of the engine. In the case of "Run unit tests", the attribute "Grade" was used to indicate the percentage of unit tests that were run without errors.

The rule shown in Table 2 as an example was actually taken from the case study, that is, it is a real gamification rule used by SWComp. We do not show the details of all the rules since they do not add much to the description of the case study.

### 5.5.2 Analysis and design of the gamified environment

One of the challenging aspects of the case study was the integration of the different tools of the company with the gamification engine, since it includes custom developments, along with COTS tools, such as TestLink, Redmine, and JUnit. Of these last three tools, JUnit presents an even more special case, since TestLink and Redmine are tools that run continuously, while JUnit is run on demand.

Figure 10 shows a diagram with the architecture of the gamified environment. There are two central elements in it: SC-Manage and the gamification engine. Since SC-Manage is a custom development, it was easy to modify this software to communicate directly with the engine. An integration component was developed, and used to carry out the communication of those behaviors related to project management and requirements management.

As regards TestLink and Redmine, there were two design choices. Because they are both open source tools, they could be modified to communicate directly with the engine. However, they also provide APIs that allow the information they manage to be reached. In the case study the second choice was preferred. As we can see in the diagram, SC-Manage integrates the information managed by TestLink and Redmine, and communicates it to the gamification engine when it detects that some of the behaviors considered have happened.

The case of JUnit was trickier since, as we have mentioned, this tool is run on demand, and does not store the results of its executions in a database. In order to integrate this tool with the engine, a wrapper was developed on JUnit. This wrapper runs the unit tests, gets the results, and communicates them to the engine.

### 5.6 Analysis of results from the case study

In this section, we analyze the results and conclusions

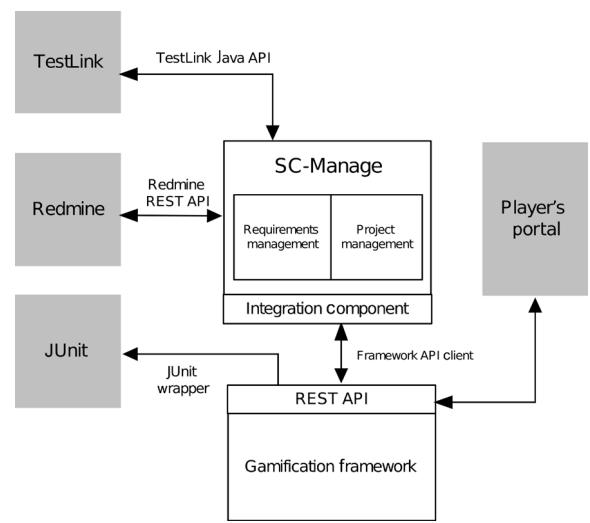

Fig. 10 Design of the gamified environment.

we can extract from the case study, following the secondary research questions of the case study.

### 5.6.1 Tool integration

Although the case study focused on the case of one single company, the tool suite we considered presented a representative example of what we can find in most software development organizations. This suite mixes custom developments with COTS tools, and it presented a case as particular as the gamification of JUnit.

The conclusion we extract from the execution of the case study is that the REST API provided by the engine is feasible for integrating the variety of tools that can be used in a real company. As regards the engine, its REST API does not force them to use any particular technology. With respect to the SE tools, most of them also provide some API that allows us to access their information. Even if they did not provide such an API, we could develop a mediator that would obtain their information by directly accessing their databases.

So far, we have focused on the integration of gamified tools with the gamification engine for the purpose of data gathering in the engine, mainly because SWComp made the decision of showing all gamification data to employee in a central player's site. However, gamification data could also flow in the opposite direction, that is, the tools could get data from the engine (such as results, rewards, etc.) to show them directly to the players. For example, this would have been interesting in the project management tool, or in the development IDE.

### 5.6.2 Support of gamification

Regarding the second research question, from this case study we conclude that the gamification abstraction on which the engine is based would support the gamification mechanisms of most companies. Actually, in the case of SC, the behaviors and rules defined did not even require all the features provided by the engine.

### 5.6.3 Integration effort

The integration of the SE tools into the gamified environment proceeded following the design and development of the engine we have just presented. The integration took 141.5 work hours. Figure 11 shows the distribution of this effort in the three areas considered in the case study. Project management was the first one to be developed, and therefore needed a greater effort because it includes the integration component of SC-Manage. Once that component was developed, the effort for integrating the other areas was significantly smaller. The effort required by the testing area (which includes JUnit, Redmine, and TestLink) is greater than that required by the requirements management area, because of the integration of SC-Manage with the testing tools.

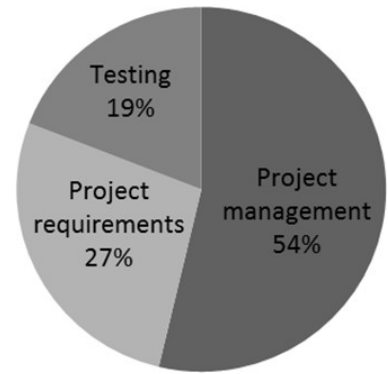

**Fig. 11 Distribution of development efforts.**

From these results, and although we cannot estimate the effort of gamifying those tools separately, we can conclude that the effort required to integrate the ecosystem of SE tools into the gamified environment was low. We could also point out that, once these tools have been integrated with the gamification engine, it would be easy to add new tools to the gamified environment, since the integration component of SC-Manage is already developed and complete.

### 5.7 Validity threats and limitations of the case study

In order to address potential threats to the validity of the case study, we considered the following:

• Construct validity: Before starting the execution of the business case, training sessions were held with employees of SC in charge of the design and development of the gamified environment, to avoid misunderstandings about the goals or scope of the project, or about the functioning of the gamification engine.

• Internal validity: To avoid other factors affecting the results of the case study, the training sessions included general knowledge about gamification; the authors of this work participated in the execution of the case study, providing support both in gamification design and in the use and configuration of the gamification engine.

• External validity: Although the results might be different in other companies, we have chosen an organization with a typical and varied tool suite that could prove how the gamification engine can support most tools present in software development companies.

● Reliability: The use of the gamification engine depends on its technical design and features, so its application in other settings should not be affected by the particular researcher applying it.

## 6　Discussion

The software architecture for gamification we present in this paper provides a valuable tool for incorporating gamification in SE workplaces composed of many tools that support different software process areas. As we explained in the previous section, the goal of the case study was to validate if our proposal is suitable for that purpose, that is, if it is able to support a wide range of tools, if the game elements it provides are able to support the gamification mechanics usually applied in software engineering, and if the gamification of a software organization's workplace can be done at a reasonable cost.

As we have seen in the presentation of the case study, the company in which we conducted it makes use of well-known off-the-shelf tools, such as Redmine, TestLink, and JUnit, and also custom developed tools. The integration of these tools into the gamification engine was not only possible but easy in all cases. Moreover, SWComp decided to develop a single interface for its employees to see the results of their actions in the gamification environment (a web called the player's portal), but these results could have also been integrated into work tools. For example, a development IDE, such as Eclipse, could have been integrated with our gamification engine, but that same IDE could also show gamification results live to the developers, since our engine not only gathers data about the tasks being completed, but it also responds with the results of evaluating those tasks and allows any tool to access all the information it manages.

The gamification elements provided by the gamification engine cover most of the general-purpose gamification elements. That is, it allows the company to implement a direct reward system in the form of points and badges. But this reward system is combined with the engine's social network to implement other gamification mechanics, such as levels and leaderboards. These gamification elements also allow us to incorporate other game mechanics, such as quests in which players can challenge other players, and even themselves. In addition, it serves as a basis for a continuous feedback system, since the player's portal shows SWComp employees real-time information on how the company is evaluating the performance they obtained in each completed task. In addition, the gamification engine provides advanced gamification mechanics, such as the virtual assistant.

The effort (and therefore the cost) of gamifying a work environment should not be forgotten due to its importance for real organizations. As we have seen in the results of the case study we conducted, the effort was really small, especially if we compare that effort with the effort of developing a custom-gamified tool for just one of the process areas we considered.

Although not initially posed as a research question of the case study, other important result can be extracted from the case study we have presented. The design of behaviors, achievements, and gamification rules provided by the framework makes the gamified work environment of SWComp very flexible. Since all the gamification logic is captured by the gamification rules, any change to the game mechanics would only require a modification of those rules through the designer's web interface, without needing to touch a line of code either on the CASE tools or in the gamification framework. We think this is a very valuable characteristic of our solution.

In addition to the core gamification aspects provided by the framework, the additional analysis functionalities it provides can be very useful. The social network analysis gives us an insight on how the players relate with each other, the weight of their relationships, and the existence of clearly defined communities. The sentiment analysis module allows us to detect problems in the motivation and happiness of the players from the messages they introduce in the system, or simply negative sentiments towards the gamified environment we have designed.

## 7　Conclusion and Future Work

In this paper, we have presented a software architecture, a gamification model, and a gamification engine for the gamification of software engineering environments. The main feature of our proposal is centralizing the logging of the behaviors of the people taking part in that environment, as well as the definition of the game rules that evaluate those behaviors and assign the corresponding achievements to them. All the business logic related to gamification is thus centralized in our engine. This allows any organization to gamify its tool suite by using the gamification engine and by carrying out easy modifications of their tools. This is an important difference compared to previous proposals, which forced the organization to either replace some of their tools with gamified tools for the same purpose, or to modify their current tools to integrate gamification into them.

The gamification is simple and general, so it can fit the work environments of most software development companies. In addition, it is easily extensible, that is, the model can accommodate any other needed type of behavior, achievement, or game rule. Another of the main benefits of using our proposals for the gamification of the tool suite of a company is that it allows us to integrate all the tools of that suite into a centralized and integrated gamified environment; that is, the rewards obtained in any of those tools add up to one total sum. An important difference our proposal has in comparison to previous approaches in gamification in SE is that it does not force the organization to replace its tools with custom-developed gamified tools.

In addition to proposing a software architecture and a gamification model from an abstract point of view, we have implemented a real gamification engine based on them. The engine not only supports the basic elements of the architecture and gamification model, but it provides advanced functionalities for gamification, such as the analysis of the interaction network derived from the collaboration of the users, which allows us to identify hub users and communities, for example. It facilitates the sentiment analysis of the texts, which can let us identify positive and negative trends in the texts the employees produce. It gives personalization support, permitting us to customize contents and functionalities in terms of the user's profile and evolution in the gamified engine, and it supplies a virtual assistant that will provide the users with help in an interactive way, using natural language, as happens in videogames, for example.

In the paper we have also presented a case study on the application of our proposal in a real organization, gamifying its whole tool suite, which includes tools common to many software development companies, such as Redmine, TestLink, or JUnit. Although we have presented our engine as a tool for the gamification of SE environments, it could easily be used in the gamification of software tools in different domains. The only limitation we find to this is given by the behavior classes currently supported by the engine. It would nonetheless be very easy to extend the behavior classes supported so that they fit new application domains.

With regard to future work, further developments are planned. Firstly, as the database of the engine contains a detailed log of all the actions carried out by developers in the SE environment, these data could be the basis for an analysis tool which extracts relevant information about the actions of the users, as well as about their performance. The engine might also be extended with a visualization component to show, for instance, user performance and rankings. Appropriate visualizations metaphors could be used (such as, for example, fish tanks with different fish species according to users' performance rates).

**Acknowledgment**

This work was supported by: For Felix Garcia and Mario Piattini: BIZDEVOPS-Global (RTI2018-098309-B-C31), Ministerio de Economía, Industria y Competitividad (MINECO) and Fondo Europeo de Desarrollo Regional (FEDER); G3Soft (SBPLY/17/180501/000150, Model Engineering for Government and Management of Global Software Development) and GEMA (Generation and Evaluation of Models for dAta Quality), Consejería de Educación y Ciencia, Junta de Comunidades de Castilla-La Mancha. For Oscar Pedreira, Alejandro Cortiñas, and Ana Cerdeira-Pena: BIZDEVOPS-Global (RTI2018-098309-B-C32), Ministerio de Economía, Industria y Competitividad (MINECO) and Fondo Europeo de Desarrollo Regional (FEDER); Datos 4.0 (TIN2016-78011-c4-1-R) and ETOME-RDF3D3 (TIN2015-69951-R), Ministerio de Economía, Industria y Competitividad (MINECO) and Fondo Europeo de Desarrollo Regional (FEDER); and Centros singulares de investigación de Galicia (ED431G/01), Grupo de Referencia Competitiva (ED431C 2017/58), and ConectaPEME GEMA (IN852A 2018/14), Xunta de Galicia y Fondo Europeo de Desarrollo Regional (FEDER).

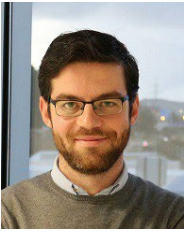
**Oscar Pedreira** is an associate professor at Universidade da Coruña (UDC), Spain. He received the MS (2006) and PhD (2009) degrees from UDC. He is a member of the Database Lab. Research group, and his research interests include data management, software engineering, and information systems.

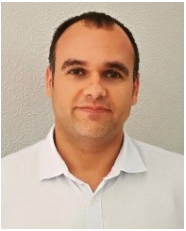
**Félix García** is a full professor at the University of Castilla-La Mancha (UCLM). He received the MS (2001) and PhD (2004) degrees from the UCLM. He is a member of the Alarcos Research Group, and his research interests include business process management, software processes, software measurement, and agile methods.

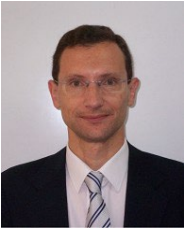
**Mario Piattini** is the director of the Alarcos Research Group and a full professor at the University of Castilla-La Mancha. He received the MS and PhD degrees from Madrid Technical University in 1989 and 1994, respectively. His research interests include information systems quality and software and data engineering.

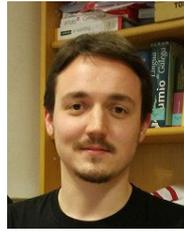
**Alejandro Cortiñas** is an assistant professor at the Database Lab of the Universidade da Coruña (Spain). He received the PhD degree from the same university in 2017 for his thesis, entitled "Software product line for web-based geographic information systems". His research topics of interest include software product lines, generative programming, geographic information systems, and spatial big data.

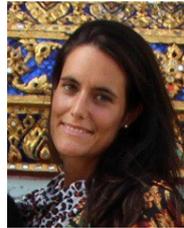
**Ana Cerdeira-Pena** obtained the MS and PhD degrees from University of A Coruña in 2007 and 2013, respectively, where she is an assistant professor. Her fields of interest include the analysis and design of compact data structures and algorithms for data compression and indexing, mathematical modelling and algorithms design for operational research problems, and information systems management.